\documentclass[%
 amsmath,amssymb,
pra,
]{revtex4-2}

\usepackage{graphicx}
\usepackage{dcolumn}
\usepackage{bm}
\usepackage{amsmath} 
\usepackage{amsfonts} 
\usepackage{amssymb} 
\usepackage{amsthm} 
\usepackage{multirow}
\usepackage{makecell}
\usepackage{wasysym}

\DeclareMathOperator*{\argmax}{argmax}
\DeclareMathOperator*{\argmin}{argmin}

\begin{document}

\preprint{APS/123-QED}

\title{Review of speckle-tracking algorithms for x-ray phase contrast imaging:\\
low dose applications}

\author{Rafael Celestre}
\affiliation{Synchrotron SOLEIL, L'Orme des Merisiers D\`epartementale 128, Saint-Aubin - France}%

\author{Laur\`ene Qu\'enot}
\affiliation{Univ. Grenoble Alpes, INSERM, UA7 STROBE, Grenoble - France}
 
\author{Christopher Ninham}
\affiliation{Univ. Grenoble Alpes, INSERM, UA7 STROBE, Grenoble - France}

\author{Emmanuel Brun}
\email{emmanuel.brun@inserm.fr}
\affiliation{Univ. Grenoble Alpes, INSERM, UA7 STROBE, Grenoble - France}

\author{Luca Fardin}
\affiliation{Univ. Grenoble Alpes, INSERM, UA7 STROBE, Grenoble - France}

\date{\today}

\begin{abstract}
X-ray speckles have been used for a wide variety of experiments, ranging from imaging (and tomography), wavefront sensing, spatial coherence measurements all the way to x-ray photon correlation spectroscopy (XPCS) and ptychography. In the near-field regime, x-ray speckle-grains preserve shape and size under free-space propagation for a static random modulation of the illumination, which permits using them as wavefront markers. The introduction of an object in the modulated field will lead to a displacement of the speckles due to refraction. Retrieving the local displacements enables access to the gradient of the phase-shift induced by the sample. The numerical process to retrieve the phase information is not trivial and numerous algorithms have been developed in the past decade with various advantages and limitations. This review focuses on near-field x-ray speckle phase imaging in the differential mode as described previously, introducing the existing algorithms with their specifications and comparing their performances under various experimental conditions.
\end{abstract}
\maketitle


\section{Introduction}\label{sec:Intro}
Over the past decades, synchrotron x-ray phase-contrast imaging (PCI) has become a great tool for the non-destructive study of samples in fields ranging from archaeology to industry and medicine \cite{Cunningham2014, Hall2021, Westneat2008, Momose2020, Quenot2022, Ponchut2021}. 
The success of this method, over conventional radiology, is partly due to the possibility of retrieving an image with enhanced edges in samples with low absorption, and enhanced contrast between sub-regions characterised by similar absorption coefficients \cite{Endrizzi2018}.
The most basic phase contrast technique used at synchrotron light sources is propagation-based imaging (PBI) \cite{snigirev1995possibilities,Cloetens1996}. This in-line phase-imaging technique is rather simple to implement needing illumination with a sufficient degree of spatial- and/or temporal coherence, with the sample and a detector at some distance downstream \cite{Wilkins1996}. However, experimental conditions and sample restrictions for which quantitative phase information can be extracted limits its applications.

\begin{figure}[h]
\includegraphics[width=0.6\textwidth]{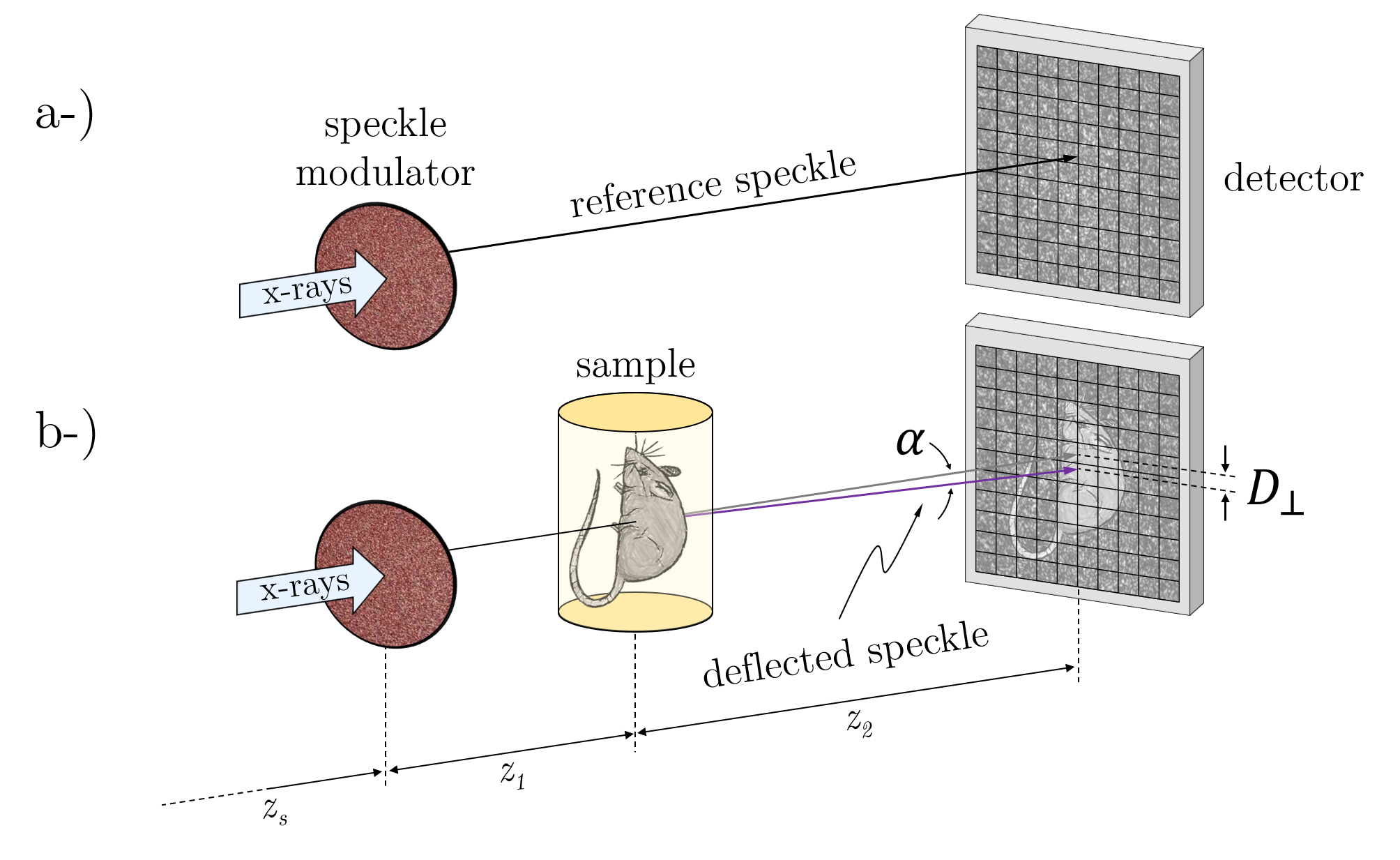}
\caption{\label{fig:set-up} Differential speckle-based imaging setup. a-) a reference image (set) is recorded with the speckle membrane in the beam; b-) the sample is then placed in the beam, distorting the speckle field. A new image (set) is recorded. The object-induced phase can be numerically retrieved by comparing the sample and the reference images. Here $z_0$ is the distance from the source to the speckle modulator, $z_1$ from the membrane until the sample and $z_2$, the sample-to-detector distance. $D_\perp$ is the transverse speckle displacement in the detector plane and $\alpha$ the associated deflection angle.}
\end{figure}

Several other PCI techniques also exist: interferometric methods (eg. Bonse-Hart interferometer); variations of the aforementioned free-space propagation technique; crystal analyser diffractometry; diffraction grating based imaging; Shack-Hartmann type sensors and other forms of wavefront markers tracking, which include speckle tracking. A more complete overview on x-ray phase-contrast imaging techniques is given by \textcite{Endrizzi2018} and \citeauthor{Weitkamp2011} -- \S~1~in~\cite{Weitkamp2011}. Speckle-based imaging (SBI) -- the topic of this work -- uses static randomly structured modulators to generate a speckle field - sand paper, filter membranes, metallic powders and coded masks are often used for that purpose \footnote{Based on accumulated experience within STROBE and the ESRF \& from private communications with X. Shi (APS/ANL) and M-C. Zdora (PSI).}. A very important characteristic of those speckle grains is that in the near-field regime they preserve shape and size, which permits their use as wavefront markers \cite{Cerbino2008, Siano2021}. In the differential implementation - see \S~2.1~in~\cite{berujon2020x} - a sample is introduced in the beam disturbing the reference speckle pattern in the detector. This is represented in Fig.~\ref{fig:set-up}. The object-induced phase can be numerically retrieved by comparing the sample and the reference images. Compared to other techniques, speckle tracking has the advantage of a very simple set-up and transfers the experimental complexity to the numerical phase extraction.

\newpage

In this article, we will present in detail a non-exhaustive list of phase retrieval algorithms for SBI. The algorithms can be sorted into two main categories: explicit tracking based on surveying the local speckle displacements through error minimisation or cross-correlation; and implicit tracking based on solving an inverse problem defined by the transport of intensity equation in each pixel. An emergent third group -- artificial intelligence-assisted methods \cite{Qiao2022} -- will not be considered in this work, because at the time of writing they are still incipient. When discussing the several implementations, we intentionally leave out the dark field calculations, as they are outside the scope of this work. The algorithms are compared under various experimental conditions. Synchrotron and laboratory sources are used to test the results under different degrees of coherence of the illumination. Samples include single material lenses and phantoms, to test for the retrieval of quantitative information, and complex biomedical tissues. For biomedical samples, the results are compared when decreasing the radiation dose, measured as the number of images required for the phase-retrieval. Reducing the radiation dose, while preserving image quality, is required to minimise the radiation damage induced on the sample and is therefore of the uttermost importance when foreseeing a clinical translation of the technique. 

\section{Phase retrieval algorithms}\label{sec:PRA}

\subsection{Explicit tracking}\label{sec:explicit_tracking}

Explicit tracking methods consist in following the local transverse displacement of the speckle patterns in the detection plane after the sample is inserted. The general idea being that a small subset of pixels centred around an arbitrary pixel $(x_i, y_j)$ in the sample image(s) and a second set of of pixels in the reference image(s) will be compared (eg. through cross-correlation), calculating the lateral shift $D_\perp $ from the coordinates $(x_i, y_j)$ that leads to the best matching between the two sub-sets. This operation is repeated for every pixel $(x,y)$ of the sample image. Among those algorithms there is a ``family" of methods that was derived from the initial x-ray speckle tracking algorithm (XST) \cite{berujon2012} and another one based on a different method called unified modulated pattern analysis (UMPA) \cite{Zanette2014}. Explicit speckle tracking methods have been extensively discussed in \cite{zdora2018state, berujon2020x}.

\subsubsection{X-ray Speckle Tracking (XST)}\label{sec:XST}

The earliest SBI algorithm, called simply x-ray speckle tracking (XST), was first published in 2012 \cite{berujon2012, morgan2012x}, and is a generalisation of a method developed for attenuation grid imaging in 2011 \cite{morgan2011quantitative}. XST requires only one pair of sample/reference images, which makes this technique particularly interesting for fast/low dose applications \footnote{In the differential implementation, calling the XST technique ``single shot'' is a bit misleading: while it is indeed true that just one image for the sample is necessary, a reference image is always needed; leading to two exposures and not a ``single shot''.}.
The method involves calculating the 2D cross-correlation between two small windows $w$ centred around the pixels $(x_i, y_j)$ and $(x_i+\chi,y_j+\gamma)$ in the sample ($I_s$) and reference ($I_r$) image respectively. $(\chi,\gamma)$ is a vector in a 2D search interval $W$ centred around $(0,0)$ that is $(2M+1)$ by $(2N+1)$ pixels wide, where $M$ and $N$ are the largest horizontal and vertical displacements in pixel units to be considered.
The transverse displacement $D_\perp(x_i,y_j)$ of the speckle modulations is given by the parameters $\chi(x_i,y_j), \gamma(x_i,y_j) \in W$ which maximise the zero-normalised cross-correlation between the two windows:
\begin{widetext}
\begin{eqnarray}
\label{eq:XST}
D_{\perp}(x_i,y_j)= \argmax_{(\chi,\gamma) ~\in~ W} \iint_{(x,y)~\in~w(x_i,y_j)}{ S(x,y)\cdot R(x+\chi,y+\gamma)~\text{d}x\text{d}y},
\end{eqnarray}
\end{widetext}
where $S= (I_s-\Bar{I_s})\big/\sigma_{I_s}$ and $R= (I_r-\bar{I_r})\big/\sigma_{I_r}$ with the bar representing the mean within the window $w$ and $\sigma$ the standard deviation. This template matching technique is analogous to the 2D Pearson product-moment correlation coefficient. The procedure described in Eq.~\ref{eq:XST} is applied to each pixel $(x_i, y_j)$ within the image or a region of interest (ROI) resulting in 2D maps $\chi(x,y)$ and $\gamma(x,y)$ of the horizontal and vertical displacements respectively. The deflection angles $\alpha = (\alpha_x,\alpha_y)$ and the transverse displacement maps $D_\perp(x, y)=(\chi,\gamma)$ are related geometrically by:
\begin{equation}\label{eq:geom_phase}
    k\alpha = k\frac{D_\perp}{z_2} = \nabla_\perp\phi,
\end{equation}
where $k$ is the wavenumber and $\nabla_\perp\phi=\Big( \partial\phi\big/\partial x, \partial\phi\big/\partial y\Big)$ is the phase transverse gradient. The phase image $\phi(x,y)$ can be obtained by numerical integration of the gradients - see \S\ref{sec:integration} for more details. To further increase the angular sensitivity of the technique, several algorithms can be used for subpixel peak detection \cite{Fisher1996,Sun2002}. Despite the procedures in Eqs.~\ref{eq:XST} and \ref{eq:geom_phase} being repeated for each pixel within a ROI containing the sample, this technique has a low spatial resolution when compared to more complex SBI methods.

\subsubsection{X-ray Speckle Vector Tracking (XSVT)}\label{sec:XSVT}

The x-ray speckle vector tracking (XSVT) method is a technique which increases the lateral resolution of XST down to a pixel and improves its angular resolution \cite{berujon2015near,berujon2016x}. 
XSVT is a scanning technique that consists of taking $\text{P}$ image pairs (sample/reference), each pair at a different (randomly chosen) transverse position $p$ of the speckle generator. The collected images are organised in 3D stacks, where the index in the third dimension corresponds to the $p^\text{th}$ membrane position. A vector is created from the sample images by reading the intensity values in a given pixel $(x_i, y_j)$ along the $\text{P}$ dimension, giving rise to the 1D signal $I_s(x_i,y_i, p)$. Following the same procedure, a set of 1D vectors is then generated from the reference images by reading the intensity values $I_r(x_i+\chi,y_j+\gamma,p)$ along P. $(\chi,\gamma)$ is a translation vector in a search interval $W$, as in XST.   

The 1D vector $I_s(x_i,y_j,p)$ is correlated with each 1D vector $I_r(x_i+\chi,y_j+\gamma,p)$, thus creating
a 2D matrix of correlation coefficients. The transverse displacement $D_\perp(x_i,y_j)$ of the speckle vector is retrieved from the coordinate $(\gamma,\chi)$ at which the 2D map of cross-correlation coefficients attains its maximum:
\begin{widetext}
\begin{eqnarray}
\begin{aligned}
\label{eq:XSVT}
D_{\perp}(x_i,y_j)= \argmax_{(\chi,\gamma) ~\in~ W} \sum_{p = 1}^{P} S(x_i,y_j, p)\cdot R(x_i+\chi,y_j+\gamma,p).
\end{aligned}
\end{eqnarray}
\end{widetext}
Again, $S= (I_s-\Bar{I_s})\big/\sigma_{I_s}$ and $R= (I_r-\bar{I_r})\big/\sigma_{I_r}$ with the bar representing the mean along the P direction and $\sigma$, the standard deviation. The sub-pixel treatments described in \cite{Fisher1996,Sun2002} and phase gradient retrieval from Eq.~\ref{eq:geom_phase} are also applied to the transverse displacement $D_{\perp}(x_i,y_j)$ obtained by Eq.~\ref{eq:XSVT}. As discussed by \textcite{qiao2020wavelet}, the direct implementation of Eq.~\ref{eq:XSVT} is very computer intensive and vectorising the operation by using matrix multiplication can speed up the calculations significantly - see \S2.1 ibid. It is clear that the XSVT technique benefits from a high number $\text{P}$ of image pairs, which clearly increases the dose the sample is subjected to and the computational effort in calculating Eq.~\ref{eq:XSVT}.

\subsubsection{The XST-XSVT hybrid}\label{sec:XST-XSVT}

Let's consider the same dataset as for a classical XSVT experiment. Instead of calculating the correlation between 1D vectors, a small window $w$ is considered around the pixel $(x_i,y_j)$ as in an XST analysis: the 3D vector $I_s(x,y,p)$, $(x,y) \in w$, is obtained. This 3D vector is correlated to a set of 3D vectors $I_r(x+\chi,y+\gamma,p)$, where $(x,y) \in w$ and $(\chi, \gamma)$ is a translation vector in the search interval $W$, as in XST and XSVT. In this case, Eq.~\ref{eq:XSVT} is rewritten as:
\begin{widetext}
\begin{eqnarray}
\begin{aligned}
\label{eq:XST-XSVT}
D_{\perp}(x_i,y_j)= \argmax_{(\chi,\gamma)~\in~ W} \sum_{p = 1}^{P} \iint_{(x,y) \in w(x_i,y_j)}S(x,y, p)\cdot R(x+\chi,y+\gamma,p)~\text{d}x\text{d}y\text{d}p
\end{aligned}
\end{eqnarray}
\end{widetext}
The window $w$ is usually kept much smaller than the interval $W$ for computational reasons. The same post-treatment dispensed on the XSVT can be applied to the lateral displacement (Eq.~\ref{eq:XST-XSVT}). This method, called XST-XSVT, was first presented in \cite{berujon2016x} as a way of reducing the number of exposures $\text{P}$ of XSVT for imaging applications, while keeping a high angular accuracy. The cost is a small decrease in lateral resolution.

\subsubsection{Wavelet X-ray Speckle Vector Tracking (WXSVT)}\label{sec:WXSVT}

The wavelet-transform-based speckle vector tracking method for x-ray phase imaging (WXSVT) was proposed by \textcite{qiao2020wavelet} as a way of improving noise robustness and increasing the computational efficiency of the aforementioned XSVT method \footnote{\textcite{Qiao2020b} have also produced a wavelet variation of the classical XST algorithm, which is not covered in this work}. The data collection is essentially the same as for the XSVT, so are the sample and reference vectors generation, that is, the 1D signal $I_s(x_i,y_i, p)$ and the 3D data set $I_r(x,y,p)$, which is limited by the sampling window $w$. 

The WXSVT method relies on applying the discrete wavelet transform (DWT) to both the sample and reference signals, re-writing them in terms of DWT coefficients. The DWT is an orthogonal transform, meaning it preserves the Euclidean distance between the sample $I_s(x_i,y_i, p)$ and the reference $I_r(x_m,y_n,p)$ vectors before and after the transformation \footnote{The Euclidean distance calculation is also an operation that can be vectorised, which is beneficial to the computational cost of the calculation.}. A property of particular interest is that the DWT allows setting cut-off coefficients enabling the reduction of the size of the transformed vector, which behave conceptually similarly to the cut-off frequencies in the Fourier transform. Up to a limit, reducing the number of detail coefficients not only improves computation efficiency, but also increases the noise robustness of the method. Alternatively, being more conservative when setting the cut-off coefficients, one can reduce the number of $\text{P}$ image pairs in the scan while still retaining good phase retrieval as demonstrated in \S3.4 from \cite{qiao2020wavelet}. This is obviously positive, from the point of view of the dose reduction in the sample. Sub-pixel maxima tracking methods used by XST and XSVT are also applied in the WXSVT method. Once the transverse displacement $D_\perp(x_i,y_j)$ is obtained, Eq.~\ref{eq:geom_phase} can be applied to obtain the phase-gradients.

\subsubsection{X-ray Speckle Scanning and its variations (XSS)}\label{sec:XSS}

X-ray speckle scanning (XSS) \cite{berujon2012x} and all its variations -- see \cite{zdora2018state,berujon2020x} -- were techniques introduced to achieve a very high angular resolution \footnote{XSS schemes have been reported to achieve nanoradian angular sensitivity \cite{Wang2015, Kashyap2016}} at the expense of a highly increased number $\text{P}$ of image pairs (sample/reference). These techniques require a much more sophisticated setup (long term stability and reproducibility of motors), with the speckle membrane being scanned with step sizes smaller than imaging detector pixel size. While these techniques are well adapted for at-wavelength metrology \cite{berujon2020x2}, they are ill-suited for low-dose or clinical applications. XSS methods are, then, out of the scope of this work.

\subsubsection{Unified Modulated Pattern Analysis (UMPA)}\label{sec:UMPA}

The unified modulated pattern analysis (UMPA) is a data processing scheme that can be applied to data sets compatible with the (W- or XST-)XSVT analysis \cite{Zanette2014, Zdora2017}. This method is based on modelling the x-ray interactions with the sample into three distinctive effects: i-) absorption in the sample reducing the intensity transmission $\mathcal{T}$; ii-) refraction causing the distortion of the speckle field $(\chi,\gamma)$; and iii-) scattering (dark field) of unresolved features in the sample decreasing the visibility of the speckle fields $\mathcal{D}$ \cite{Zanette2014} - see also supplementary material ibid. The x-ray beam intensity in the presence of the sample at the detector plane can be modelled as:
\begin{eqnarray}
\begin{aligned}
\label{eq:UMPA_intensity}
I(x,y) = \mathcal{T}\cdot\big[ I_0(x-\chi,y-\gamma) + \mathcal{D}\cdot\Delta I_r(x-\chi,y-\gamma)\big] \quad 
\end{aligned}
\end{eqnarray}
where $I_r(x,y)= I_0(x,y) +  \Delta I_r(x,y)$ is the reference signal decomposed into an average-valued signal $I_0(x,y)$ and the fluctuations around this DC value $\Delta I_r(x,y)$. The UMPA method obtains the signals $\mathcal{T}$,  $\mathcal{D}$ and the displacement maps $(\chi,\gamma)$ by minimising the sum of squared differences of the cost function:
\begin{widetext}
\begin{eqnarray}
\begin{aligned}
\label{eq:UMPA_loss}
\mathcal{L}(x_i,y_i&; \chi, \gamma, \mathcal{T}, \mathcal{D})  \\&=\sum_{p = 1}^{P} \iint_{x,y~\in~ w(x_i,y_j)} \Big\{\mathcal{T}(x,y,p)\cdot\big[ I_0(p) + \mathcal{D}(x,y,p)\cdot\Delta I_r(x-\chi,y-\gamma,p)\big] - I_s(x,y, p)\Big\}^2~\text{d}x\text{d}y
\end{aligned}
\end{eqnarray}
\end{widetext}
Both $\mathcal{T}$ and $\mathcal{D}$ have analytical expressions, which should be plugged into Eq.~\ref{eq:UMPA_loss} prior to the minimisation - see Eqs.~5~and~6 in the supplemental material from \cite{Zanette2014}. Much like the XST-XSVT method, the signals $I_r$ and $I_s$ are also limited by the window functions $w$ and $W$ centred around the pixel $(x_i,y_i)$. 
\begin{eqnarray}
\begin{aligned}
\label{eq:UMPA}
D_{\perp}(x_i,y_i),~\mathcal{T}(x_i,&y_i),~\mathcal{D}(x_i,y_i) = \\ &\argmin_{\chi,\gamma,\mathcal{T}, \mathcal{D}}~\mathcal{L}(x_i,y_i; \chi, \gamma, \mathcal{T}, \mathcal{D})
\end{aligned}
\end{eqnarray}
Sub-pixel algorithms are also employed for increasing the angular sensitivity in $D_\perp$ and the final conversion from transverse displacement $\chi$ and $\gamma$ into phase gradients is done using Eq.~\ref{eq:geom_phase}. A very informative overview of the UMPA numeric implementation as well as latest developments are given by \textcite{DeMarco2023}.

\subsubsection{Comments on the explicit tracking methods}\label{sec:explicit_comments}

The methods presented hitherto along with their technical challenges and algorithmic implementations are described in more details in \cite{zdora2018state,berujon2020x,berujon2020x2, qiao2020wavelet}. Despite efforts in optimisation, the main drawback of these methods is their computational cost. Without powerful computers and code parallelisation, retrieving the phase for a sample may take quite a long time depending on the number of sample/reference image pairs and search windows. Coupling these techniques with tomography, where each projection must undergo the full phase-retrieval pipeline, can become impractically long, for example.

\subsection{Implicit tracking}\label{sec:implicitt_tracking}

Implicit speckle tracking is the general name given to a family of techniques based on solving the transport of intensity equation (TIE) under different assumptions related to the sample. For a monochromatic scalar paraxial electromagnetic wave with intensity $I(x,y,z)$ propagating along the $z-$direction (optical axis):
\begin{equation}
\label{eq:TIE}
\frac{\partial}{\partial z}I(x,y,z) =-\frac{1}{k}\nabla_{\perp}\big[I(x,y,z)\cdot\nabla_{\perp} \phi(x,y,z)\big]
\end{equation}
Eq.~\ref{eq:TIE} (TIE) can be derived from the paraxial Helmholtz equation, meaning it is subjected to the same assumptions used in the scalar diffraction theory and the paraxial approximation in optics \cite{zuo2020transport, Paganin2006}; furthermore, it is relevant to mention that the TIE is an expression of energy conservation, relating the axial intensity derivative (left-handed side) with the total energy variation in the transverse plane (right-handed side) \cite{zuo2020transport}. Eq.~\ref{eq:TIE} can be approximated by its finite-difference form for an x-ray beam flowing through a sufficiently small distance (i.e. high Fresnel number) $d_z$:
\begin{eqnarray}
\begin{aligned}
\label{eq:TIE_finitediff}
I(x,y,z+d_z)-I(x,y,z&)\approx\\-\frac{\textit{d}_z}{k}\nabla_{\perp}\big[&I(x,y,z)\cdot\nabla_{\perp}\phi(x,y,z)\big]
\end{aligned}
\end{eqnarray}
Which is the form of the TIE used for deriving the implicit methods presented here. A very complete tutorial on the TIE and its applications together with several methods for deriving Eq.~\ref{eq:TIE} and algorithms for solving it are presented by \textcite{zuo2020transport}. A generalisation of the TIE for partially-coherent x-ray paraxial imaging and applications to phase imaging -- the Fokker-Planck equation (F-PE) -- has been proposed by \textcite{Paganin2019} as a way of accounting for diffusive paraxial energy transport -- similar to what is conveyed by $\mathcal{D}$ in the UMPA method in \S\ref{sec:UMPA}.

\subsubsection{Optical Flow algorithm (OF)}\label{sec:OF}

\label{eq:TIESpeckle}
In 2018, \textcite{paganin2018single} proposed applying the finite difference version of the TIE (Eq.~\ref{eq:TIE_finitediff}) adapted to the SBI problem assuming a perfectly transparent sample (ideal phase element) distorting the speckle field \footnote{Applying the chain-rule on the right-handed side of Eq.~\ref{eq:TIE_OF}, one obtains two terms: $\big[\nabla_{\perp}I_r(x,y)\cdot\nabla_{\perp}\phi(x,y)+ I_r(x,y)\cdot\nabla^2_{\perp}\phi(x,y)\big]\cdot z_2\big/k$. The first term is often referred to as the ``prism'' term and is responsible for laterally shifting the image at $z+d_z$, while the second term is called ``lensing-term'' (de-)concentrating light. The lensing-term (or Laplacian term) corresponds to the variations of intensity known as phase-contrast.} under a quasi-coherent illumination \cite{Gureyev2006}:
 
\begin{equation}
\label{eq:TIE_OF}
I_{r}(x,y)-I_{s}(x,y) \approx \frac{z_2}{k} \nabla_{\perp}[I_{r}(x,y)\cdot\nabla_{\perp}\phi(x,y)]
\end{equation}
The term $\nabla_{\perp}\phi(x,y)$ in Eq.~\ref{eq:TIE_OF} can be replaced by the relation expressed in Eq.~\ref{eq:geom_phase}, resulting in:
\begin{equation}
\label{eq:TIE_OF_simple}
I_{r}(x,y)-I_{s}(x,y) \approx \nabla_{\perp}[I_{r}(x,y)\cdot D_{\perp}(x,y)]
\end{equation}
It is important to recall that $D_{\perp}(x,y)$ is of vectorial nature and is composed of the scalar fields $\chi(x,y)$ in the $x-$direction and $\gamma(x,y)$ in the $y-$direction. In order to facilitate solving Eq.~\ref{eq:TIE_OF_simple}, the multiplicative term $I_{r}(x,y)\cdot D_{\perp}(x,y)$ is approximated by the gradient of an auxiliary scalar function $\nabla_\perp \Lambda(x,y)$ -- see \cite{paganin2018single} for entailing approximations leading to Eq.~3 ibid. This manipulation allows Eq.~\ref{eq:TIE_OF_simple} to be rewritten as a Poisson-type equation:
\begin{equation}
\label{eq:TIE_OF_Poisson}
I_{r}(x,y)-I_{s}(x,y) \approx \nabla^2_{\perp}\Lambda(x,y) 
\end{equation}
A simple way to solve Eq.~\ref{eq:TIE_OF_Poisson} and extract the lateral displacements $D_{\perp}(x,y)$ is by using Fourier direct and inverse transforms ($\mathcal{F}$ and $\mathcal{F}^{-1}$) as follows:
\begin{eqnarray}
\begin{aligned}
&{D}_{\perp}(x,y) = \\ &\frac{i}{I_r(x,y)}\mathcal{F}^{-1} \left\{ (\kappa_x,\kappa_y)  \frac{\mathcal{F}\{I_s(x,y)-I_r(x,y)\}}{\kappa_x^2+\kappa_y^2} \right\}
\label{eq:OpticalFlowGradient}
\end{aligned}
\end{eqnarray}
where $i$ is the imaginary unity and $(\kappa_x,\kappa_y)$ are the coordinates in the Fourier space. After that, Eq.~\ref{eq:geom_phase} is used to retrieve the phase gradients.

Solving the TIE equation under the assumptions leading to Eq.~\ref{eq:OpticalFlowGradient} -- i.e. using the optical flow (OF) method --  has two main advantages: i-) numerically solving Eq.~\ref{eq:OpticalFlowGradient} is very fast and computationally efficient -- only two FFTs are necessary; and ii-) only one sample/reference pair is required, which makes the OF very dose-efficient.

However, this method assumes a non-absorbing sample, which is rarely the case for clinical samples with dense structures. The absorption due to the sample can be partially estimated and corrected for by blurring $I_s(x,y)$ and $I_r(x,y)$, to reduce the speckle modulation, and by calculating the ratio of the resulting images. A drawback of the OF method is that Eq.~\ref{eq:OpticalFlowGradient} approaches a singularity as $\kappa_x^2+\kappa_y^2\rightarrow0$. This can be tackled by masking out the singular point \cite{paganin2018single} or by applying a Gaussian-shaped high-pass filter in the signal \cite{rouge2021comparison}. Excessive filtering out lower frequencies can be detrimental depending on the final application.

\subsubsection{Single Material Object Speckle Tracking (SMOST)}\label{sec:SMOST}
The SMOST method was conceived to extend the OF algorithm to monomorphous absorbing samples \cite{pavlov2019single, Gureyev2015_mono}. It starts by manipulating Eq.~\ref{eq:TIE_OF} into:
\begin{equation}
\label{eq:TIE_Pavlov}
\frac{I_{s}(x,y)}{I_{r}(x,y)} \approx  I_\text{obj}(x,y)-\frac{z_2}{k} \nabla_{\perp}[I_\text{obj}(x,y)\cdot\nabla_{\perp}\phi_\text{obj}(x,y)]
\end{equation}
where $I_\text{obj}(x,y)$ and $\phi_\text{obj}(x,y)$ are the sample transmitted intensity and impinged phase -- refer to \cite{pavlov2019single} for a derivation of Eq.~\ref{eq:TIE_Pavlov}. This sample has projected thickness along the optical axis given by $\Delta_z(x,y)$ and is composed of a material with a complex index of refraction $n = 1 -\delta + i\cdot\beta$ \footnote{The values for $\delta$ (index of refraction decrement) and $\beta$ (absorptive part) are well documented in tables \cite{Hubbell1975,Henke1993} and computer libraries \cite{brunetti_library_2004}}. If the sample is well approximated by a thin-single-material object \cite{Paganin2006}, it induces a phase-shift $\phi_\text{obj}(x,y)=-k\delta\Delta_z(x,y)$ and an intensity transmission $I_\text{obj}(x,y)=\exp{[-2k\beta\Delta_z(x,y)]}$. Substituting these two expressions into Eq.~\ref{eq:TIE_Pavlov} leads to:
\begin{equation}
\label{eq:TIE_Pavlovbis}
\frac{I_{s}(x,y)}{I_{r}(x,y)} \approx \bigg( 1-\frac{z_2}{2k}\frac{\delta}{\beta}\nabla_\perp^2\bigg)\exp{}\big[-2k\beta \Delta_z(x,y)\big]
\end{equation}
Eq.~\ref{eq:TIE_Pavlovbis} can be solved for $\Delta_z(x,y)$ by:
\begin{eqnarray}
\begin{aligned}
\Delta_z(x,y) &=\\-\frac{1}{\mu}&\ln\Bigg(\mathcal{F}^{-1} \left\{ \frac{\mathcal{F}\{I_s(x,y)\big/I_r(x,y)\}}{1+z_2(\delta\big/\mu)(\kappa_x^2+\kappa_y^2)} \right\}\Bigg)
\label{eq:TIE_PavlovSolution}
\end{aligned}
\end{eqnarray}
where $\mu = 2k\beta$. Due to the Fourier transforms involved, filtering is sometimes employed to reduce low-frequency artefacts. Unlike the previous methods that retrieve the transverse displacement $D_\perp(x,y)$, SMOST retrieves directly the sample thickness in projection approximation $\Delta_z(x,y)$, which is linearly proportional to the phase-shift $\phi(x,y)$ for fixed energy. As pointed out by \textcite{pavlov2019single}, Eq.~\ref{eq:TIE_PavlovSolution} bares strong resemblance with \textcite{paganin2002simultaneous} method for propagation-based x-ray phase contrast imaging, which makes this method particularly attractive to low-flux sources or conversely, low-dose applications.

\subsubsection{Low Coherence System algorithm (LCS)}\label{sec:LCS}

The assumptions used in deriving the OF algorithm are quite restrictive and not always attainable for samples of interest or under common experimental conditions. Although SMOST already handles absorbing samples, it still relies on a coherent illumination - this can be seen by the Laplacian dependency of the phase in Eq.~\ref{eq:TIE_Pavlovbis}. Relaxation of that hypothesis in OF or SMOST, namely the necessity of transparent sample and high coherence, leads to the low coherence system algorithm (LCS) \cite{quenot2021towards,quenot2021optica}. To begin with, the sample image in the presence of the speckle field $I_s$ can be corrected by an additional ``loss term'' $I_\text{obj}$ to account for the attenuation due to the sample. Eq.~\ref{eq:TIE_OF_simple} can, then, be re-written as:
\begin{eqnarray}
\begin{aligned}
\label{eq:TIE_LCS_A}
I_{r}(x,y)-\frac{I_{s}(x,y)}{I_\text{obj}(x,y)} \approx \nabla_\perp\left[I_{r}(x,y)\cdot D_{\perp}(x,y)\right]
\end{aligned}
\end{eqnarray}
The manipulation in Eq.~\ref{eq:TIE_LCS_A} allows the TIE framework to be extended into absorbing samples. We now expand the right-handed side of Eq.~\ref{eq:TIE_LCS_A} using the chain rule:
\begin{eqnarray}
\begin{aligned}
\label{eq:TIE_LCS_B}
I_{r}(x,y)-\frac{I_{s}(x,y)}{I_\text{obj}(x,y)} \approx \nabla_\perp &I_r(x,y)\cdot D_\perp(x,y) &&\\ 
 + &I_r(x,y)\cdot \nabla_\perp D_\perp(x,y)
\end{aligned}
\end{eqnarray}
We recall that $ \nabla_\perp D_\perp(x,y)=z_2\big/k \cdot \nabla^2_\perp \phi(x,y)$. 
\citeauthor{quenot2021optica} claimed that, if the system has low coherence, the Laplacian term could be neglected as the the interference fringes could not be resolved \cite{quenot2021optica}. This leads to:
\begin{eqnarray}
\begin{aligned}
\label{eq:TIE_LCS}
I_{r}(x,y)-\frac{I_{s}(x,y)}{I_\text{obj}(x,y)} \approx &&\\
\chi(x,y)\frac{\partial}{\partial x}&I_r(x,y)+\gamma(x,y)\frac{\partial }{\partial y}I_r(x,y)
\end{aligned}
\end{eqnarray}
Note that $\nabla_\perp I_r$ goes to $0$ in the absence of a speckle field with high visibility and small speckle grains: in this case, no (good) phase retrieval can be performed. Eq.~\ref{eq:TIE_LCS} has three unknown variables: the two transverse displacement arrays $\chi$ and $\gamma$; and the loss term $I_\text{obj}(x,y)$, hence a system with $p \in \{1~...~P \},~\text{P}\geq 3$ equations can be solved for those three aforementioned variables:
\begin{widetext}
\begin{eqnarray}
\begin{aligned}
\label{eq:TIE_LCS_system}
I_{r}^{(p)}(x,y) = \frac{1}{I_\text{obj}(x,y)}I_{s}^{(p)}(x,y) + \chi(x,y)\frac{\partial}{\partial x}I_r^{(p)}(x,y)+\gamma(x,y)\frac{\partial }{\partial y}I_r^{(p)}(x,y)
\end{aligned}
\end{eqnarray}
\end{widetext}
where each equation with a superscript $p$ corresponds to an image pair (sample/reference), each pair at a different (randomly chosen) transverse position $p$ of the speckle generator.
Although Eq.~\ref{eq:TIE_LCS} assumes a incoherent illumination, if the sample is indeed illuminated by a coherent beam, the phase contrast fringes that arise due to the Laplacian term in Eq.~\ref{eq:TIE_LCS_B} will be attributed to the loss term $I_\text{obj}(x,y)$ when solving the system in Eq.~\ref{eq:TIE_LCS_system}. This algorithm is inherently sensitive to sub-pixel displacements \cite{quenot2021optica}. Once the vectors $\chi(x,y)$ and $\gamma$ are found, Eq.~\ref{eq:geom_phase} relates them to the phase gradients.

\subsubsection{X-ray Multi-Modal Intrinsic-Speckle-Tracking (MIST)}\label{sec:MIST}

The MIST family of methods is based on the Fokker-Planck equation, a generalisation of the TIE \cite{Paganin2019}:
\begin{eqnarray}
\begin{aligned}
\label{eq:FPE}
I_{r}(x,y)-I_{s}(x,y) \approx &\frac{z_2}{k} \nabla_{\perp}[I_{r}(x,y)\cdot\nabla_{\perp}\phi(x,y)]&&\\
-&z_2 \mathcal{D}(x,y)\cdot\nabla^2_{\perp}I_r(x,y)
\end{aligned}
\end{eqnarray}
where $\mathcal{D}(x,y)$ is the diffusive (dark field) term. Comparing the F-PE (Eq.~\ref{eq:FPE}) to the TIE (Eq.~\ref{eq:TIE_OF}) allows to say that the MIST methods are a natural expansion of the OF-based ones. Solving Eq.~\ref{eq:FPE} can be done applying the same assumptions used for SMOST for the first term in the right-hand side of the F-PE and several other assumptions for expanding the second term \cite{pavlov2020x, pavlov2021,Alloo2022,alloo2023}. This review does not cover the calculation or applications of the $\mathcal{D}$ signal and therefore the MIST family will not be considered here.

\subsection{Integration methods}\label{sec:integration}

With the exception of the SMOST method, all algorithms presented here retrieve the transverse deflection maps $D(x,y)=(\chi,\gamma)$. Eq.~\ref{eq:geom_phase} relates the deflection maps to the horizontal and vertical phase gradients:
\begin{equation}\label{eq:phase_grad}
    \frac{\partial}{\partial x}\phi(x,y) = \frac{k}{z_2}\chi(x,y); \quad \frac{\partial}{\partial y}\phi(x,y) = \frac{k}{z_2}\gamma(x,y)
\end{equation}
Phase images are obtained, by integrating numerically the gradients in Eq.~\ref{eq:phase_grad}, a problem which, in the presence of noise, is ill-posed \cite{Ettl2008,Huang2015}. Several two-dimensional integration methods for surface reconstruction from gradient data were developed by \textcite{frankot1988method}, \textcite{arnison2004linear} and \textcite{Kottler2007}. 
In this work, Frankot Chellappa's integration method (FC) is used to compare the different phase retrieval algorithms:
\begin{widetext}
\begin{eqnarray}
\begin{aligned}
\label{eq:Frankot}
\phi(x,y) = \mathcal{F}^{-1}\Bigg\{-\frac{i}{2\pi(\kappa_x^2 +\kappa_y^2)}\Bigg(\kappa_x\mathcal{F}\bigg\{\frac{\partial}{\partial x}\phi(x,y)\bigg\} + \kappa_y\mathcal{F}\bigg\{\frac{\partial}{\partial y}\phi(x,y)\bigg\} \Bigg)\Bigg\}
\end{aligned}
\end{eqnarray}
\end{widetext}
FC is a Fourier-based solver and assumes periodic boundary conditions, which are often not satisfied, especially when the sample extends beyond the edges of the field of view. To avoid artefacts in the integrated phase, an anti-symmetrization of $\chi$ and $\gamma$ was implemented, as suggested by \textcite{Bon2012}.

\subsection{Numerical implementation}\label{sec:num_imp}

All the phase retrieval algorithms presented here, together with several numerical integration methods, are conveniently centralised in an open-access Python library called POPCORN (\textbf{PO}st-processing \textbf{P}hase \textbf{CO}ntrast and spect\textbf{R}al x-ray imagi\textbf{N}g) \cite{POPCORN}. The UMPA code was taken originally from \cite{UMPA} and an updated version is available at \cite{UMPA2}. The WXSVT implementation is taken from \cite{WXSVT}.

\begin{widetext}
\begin{table*}[t]
\caption{\label{tab:Experiments_param}Samples description \& experiments parameters}
\resizebox{\textwidth}{!}{%
\begin{tabular}{ccccccccc}
\hline \hline
 \textbf{sample} & \textbf{source} &\textbf{ beam energy} [keV] &\textbf{ prop. distance} [m] &~~~~\textbf{beam modulator}~~~~& ~~\textbf{pix. size} [µm]~~& \textbf{detector} & \textbf{section} (\S)\\ \hline
 \makecell{nylon wires\\($\diameter$140~µm, $\diameter$200~µm)} & ID17 & 52.0 & 3.600 & sandpaper & 3.0 & pco.edge & \ref{sec:phantom} \\ \hline
 \makecell{x-ray lenses\\(50~µm, 500~µm, 5000~µm)} & BM05 & 20.0  & 0.750 & \makecell{filter membrane\\(pore size $\sim$1.2~µm)} & 1.6 & pco.edge& \ref{sec:metrology}\\ \hline
 mouse knee & ID17 & 52.0 &  11.000 & sandpaper & 6.1 & pco.edge & \ref{sec:mouse_knee}\\ \hline
 fly on straw&~~Easytom XL-RX~~&~~W anode at 40.0~kVp~~& 0.345 & \makecell{TiC powder\\(grain size $\sim$100~µm)} & 48.0 & Varian flat panel & \ref{sec:fly}\\ \hline \hline
\end{tabular}}
\end{table*}
\end{widetext}

\section{Methods: Samples description \& experiments parameters}\label{sec:exp_parm}

The phase retrieval schemes, previously presented, were applied to four different sample groups under diverse experimental conditions.
The relevant experimental parameters are summarised in Table~\ref{tab:Experiments_param}. 

The first samples were cylindrical nylon wires of 140~µm and 200~µm diameter, which were measured on the ID17 beamline at ESRF. Nylon wires are a common phantom for x-ray phase imaging and were chosen to benchmark quantitatively the speckle tracking algorithms as they offer a simple analytical model for the phase gradient. The second set is composed of three bi-concave parabolic x-ray lenses made out of beryllium \cite{Lengeler2004}, which are typical samples for metrology measurements. These focusing lenses have different characteristic radii (50~µm, 500~µm and 5000~µm) and were chosen due to their varying strength of phase modulation: the shorter the radius, the stronger the phase variation leading to larger speckle displacements. The lenses were measured on the BM05 beamline at ESRF. A mouse knee was chosen for a qualitative evaluation of the algorithms when applied to biomedical samples. Biomedical samples are characterised by high spatial frequencies, discontinuities and interfaces between different materials, which should be correctly retrieved while reducing the radiation dose. These measurements were also performed at the ID17 beamline. The experiments described so far were performed on a synchrotron source with monochromatic illumination achieved with a Si(111) double-crystal monochromator $(\Delta\text{E}\big/\text{E}\approx10^{-4})$. Furthermore, due to the long propagation distances between the source and speckle modulator, the illumination presents elevated degree of spatial coherence -- cf. the van-Cittert-Zernike theorem in \cite{Gureyev2017} and the references therein. Finally, the last sample -- a headless domestic fly -- was measured on a micro-focus laboratory x-ray source with a broad spectrum (tungsten anode at 40 kVp) in order to test the algorithms on yet another complex sample with small features but under a low-coherence illumination. The measurements were performed on an adapted EasyTom XL tomographic set-up from RX solutions at the SIMaP (Univ. Grenoble Alpes). 

\begin{figure*}[t]
	\includegraphics[width=0.9\textwidth]{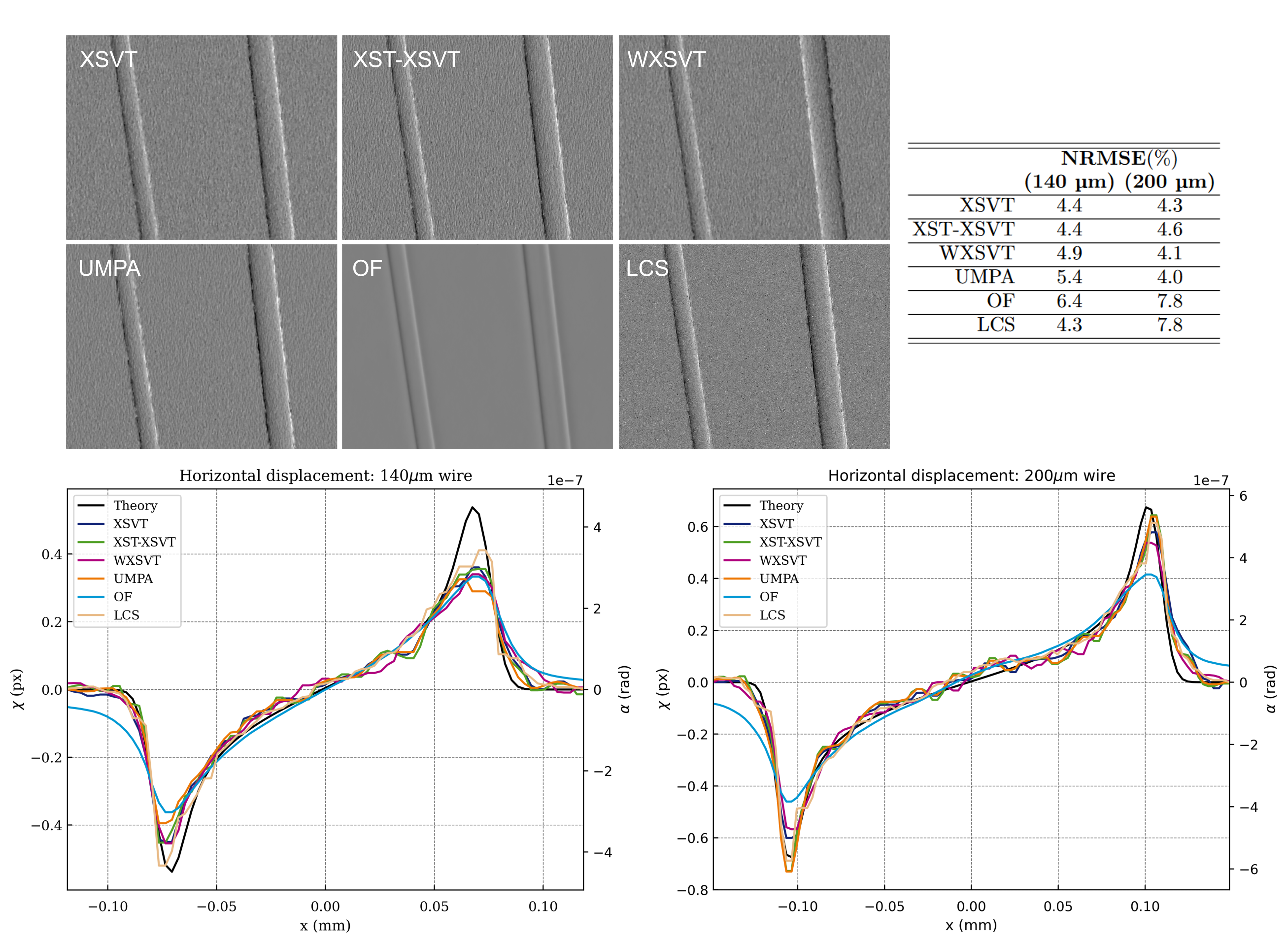}
	\caption{Nylon wires displacement maps retrieved with different algorithms (XSVT, XST-XSVT, WXSVT, UMPA, OF, LCS). Plots of the wires profiles compared to theory. Table of the NRMSE calculated between the experimental profiles and the theoretical one.}
	\label{fig:ComparisonNylonWireDx}
\end{figure*}

\section{Results: algorithms comparison}\label{sec:comp}

\subsection{Nylon wires: phantom imaging \& quantitative analysis}\label{sec:phantom}

This round-robin study was conducted on the nylon wires. The theoretical phase shift induced by a single-material cylindrical sample is known, therefore only the algorithms that retrieve displacement maps were compared: XSVT, XST-XSVT, WXSVT, UMPA, OF and LCS. The phase gradients in this section were obtained by applying the several algorithms to a set of 10 reference and 10 sample images at different membrane positions. Figure~\ref{fig:ComparisonNylonWireDx} shows horizontal displacement for two nylon wires with nominal diameter of 140~µm and 200~µm. Since they are tilted with respect to the vertical axis, the slanted edge method was adopted to obtain a perpendicular profile cut. This method is typically used for the calculation of a super-sampled Edge Spread Function (ESF) or the Modulation Transfer Function (MTF) \cite{Reichenbach91}. To reduce noise-induced fluctuations, 200 profiles were averaged per wire in Fig.~\ref{fig:ComparisonNylonWireDx}, which also shows the displacement field in units of pixel and in angular deflection (Eq.~\ref{eq:geom_phase}). A theoretical profile is also shown: it was computed from the gradient of the phase shift induced by a perfect cylindrical Nylon wire at the nominal beam energy. The Point Spread Function of the detector, here of 2 pixels \cite{Mittone_scmos}, was simulated by applying a Gaussian filter to the phase-shift gradient.

In this experiment all algorithms provided a displacement close to the theoretical values, which is confirmed by the calculation of the normalised root-mean-square error (NRMSE) - the NRMSE stays under 5.5~\% for both wires over a wide choice of algorithms. Possible sources of errors can be difficult to pinpoint, but we list some of them here: regardless of the technique, the edges of the wires are smoothed and this is mainly due to the imaging system lateral resolution (scintillator blurring and imaging system PSF). This is also where the displacement field has a discontinuity; We also bring the attention to residual propagation-based edge-enhancement effects, which depend on the Laplacian of the phase-shift; Lastly, the non-perfect cylindrical shape of the nylon wires cause difference between observed and calculated values. Optical flow provides the least accurate quantitative information which depends strongly on the high-pass filter applied, despite good qualitative results - for single-material samples, attenuation and phase shift are proportional and related to the thickness of the sample. For this reason, it cannot be excluded that the shape of the profile measured by OF was dominated by the attenuation of the wires. 

\begin{figure*}[t]
	\includegraphics[width=0.9\textwidth]{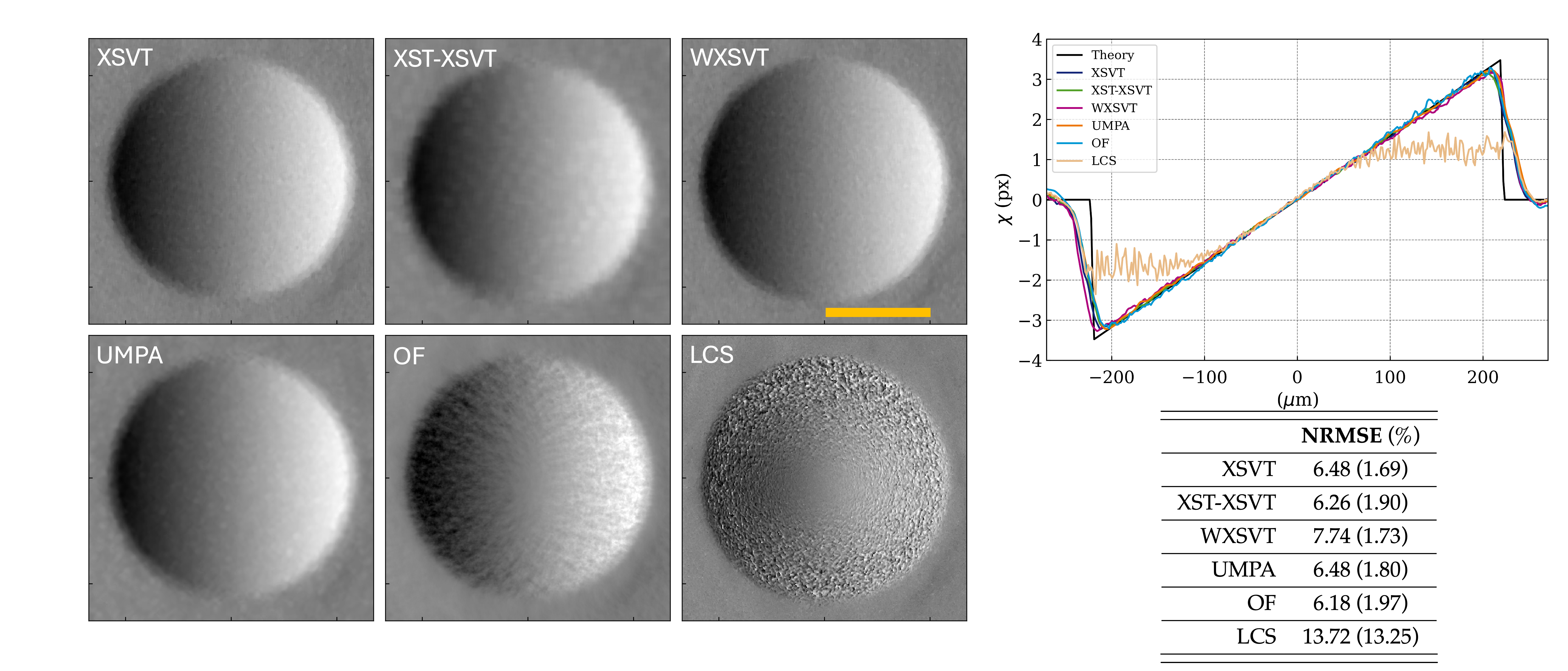}
	\caption{Displacement maps of a 2D beryllium lens with $R=50$~µm from measurements with 10 membrane positions. The NRMSE values are calculated within the limits of the profile cuts above, while values in brackets are calculated within the lens active area. The horizontal profile cuts were obtained the the centre of the lens. The orange bar represents 200~µm.}
	\label{fig:Comparison_lens_R50}
\end{figure*}

\begin{figure*}[t]
	\includegraphics[width=0.9\textwidth]{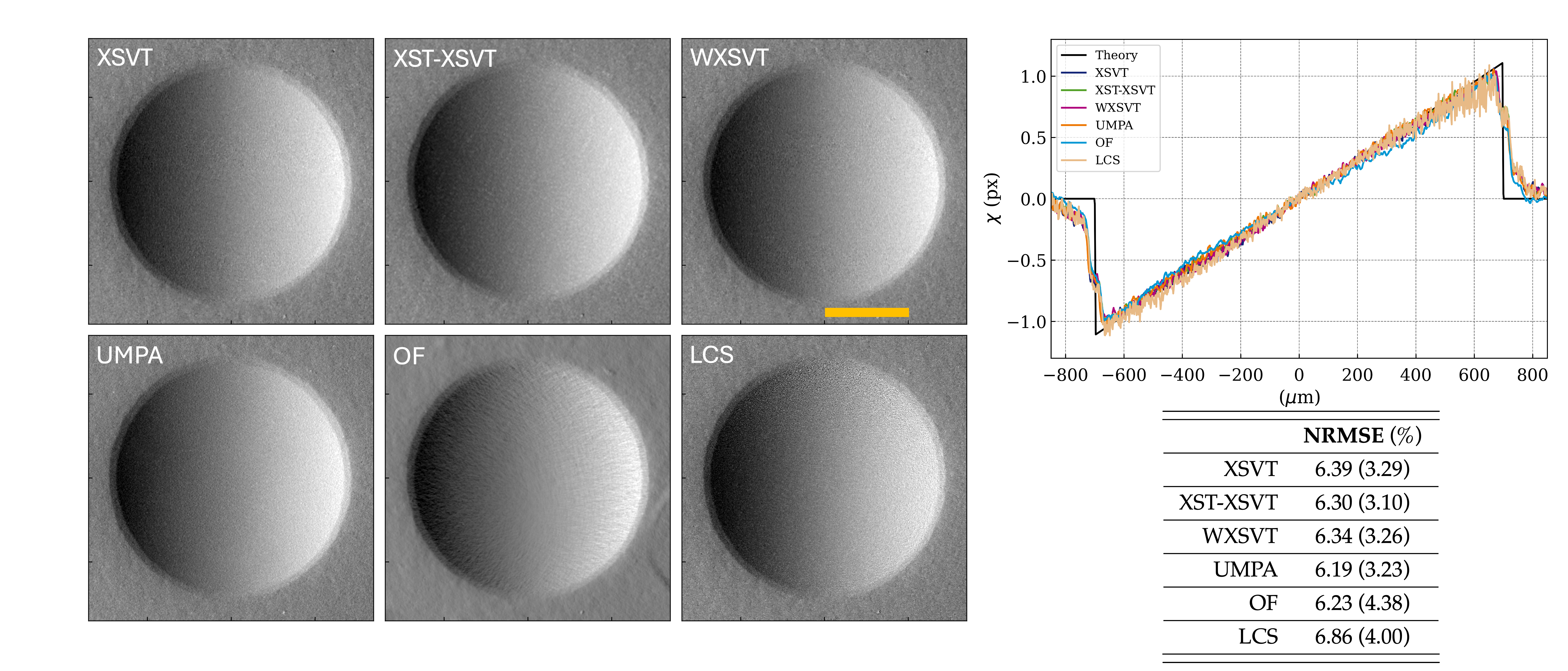}
	\caption{Displacement maps of a 2D beryllium lens with $R=500$~µm from measurements with 10 membrane positions. The NRMSE values are calculated within the limits of the profile cuts above, while values in brackets are calculated within the lens active area. The horizontal profile cuts were obtained the the centre of the lens. The orange bar represents 500~µm.}\label{fig:Comparison_lens_R500}
\end{figure*}

\begin{figure*}[]
	\includegraphics[width=0.9\textwidth]{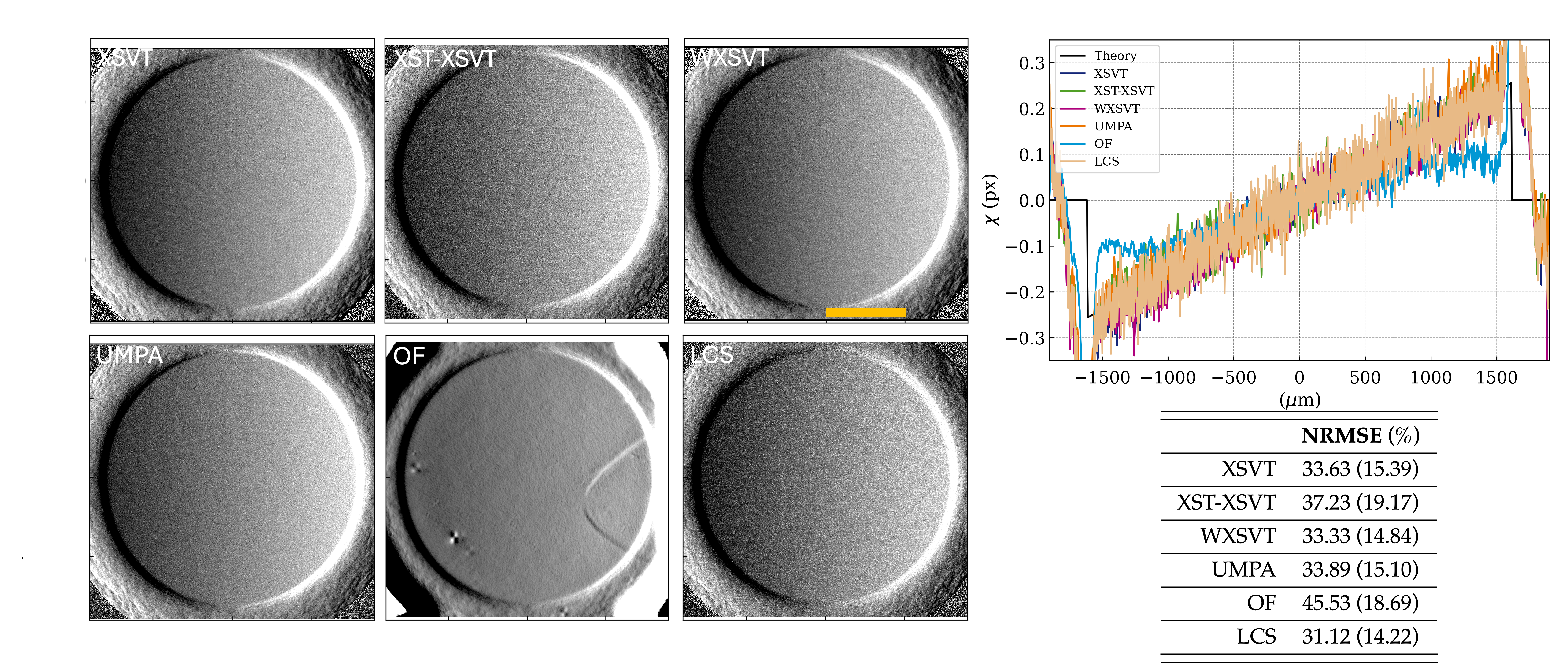}
	\caption{Displacement maps of a 2D beryllium lens with $R=5000$~µm from measurements with 10 membrane positions. The NRMSE values are calculated within the limits of the profile cuts above, while values in brackets are calculated within the lens active area. The horizontal profile cuts were obtained the the centre of the lens. The orange bar represents 1000~µm.}\label{fig:Comparison_lens_R5000}
\end{figure*}

\subsection{X-ray lenses and larger pixel displacements}\label{sec:metrology}

A second quantitative study was performed on refractive x-ray lenses with radii varying from 50~µm, 500~µm to 5000~µm - the smaller the radius, the shorter the focal length of the lens and the larger the speckle lateral displacement will be. Under the experimental conditions summarised in Table~\ref{tab:Experiments_param}, their respective focal lengths are 29.36~m, 293.58~m and 2.94~km. X-ray lenses attenuate the incoming beam, but the lenses used in this experiment are made of beryllium and at 20~keV absorption is kept to a minimum. X-ray lenses can be easily modelled and theoretical values for their phase gradients readily obtained \cite{celestre_recent_2020, Celestre:Tilted}. Furthermore, the phase gradients of a 2D parabolic lens is linear in both vertical and horizontal directions within the lens geometric aperture, which makes visual inspection of the displacement graphs in Figs.~\ref{fig:Comparison_lens_R50}-\ref{fig:Comparison_lens_R5000} straightforward. All these make x-ray lenses good phantoms for SBI and the same treatment dispensed to the nylon wires will be applied to the lenses. Much like in the previous section, the phase gradients presented here were obtained by applying the several algorithms to a set of 10 reference and 10 sample images at different speckle membrane positions.

Figure~\ref{fig:Comparison_lens_R50} shows the horizontal phase gradients and the NRMSE values for a 50~µm radius lens. For this experiment configuration, the speckle shift at the edge of the lens is expected to be slightly over 3 pixels. We can see that with the exception of the LCS algorithm - that stagnates around $\pm$~1 pixel, all algorithms follow closely a linear profile and overlap well with the theoretical curve within the lens active area, however, the edges of the lenses are not well represented by the theoretical model and hence the difference between experimental and simulated data reflected by the NRMSE - this mismatch has been documented and discussed into depth in a previous work \cite{Celestre:Tilted}. The NRMSE vales are calculated within the limits of the profile cuts in Fig.~\ref{fig:Comparison_lens_R50}, while the values in brackets are calculated within the lens active area and hence their lower values (meaning better agreement). A visual inspection of the 2D maps shows that XSVT, WXSVT and UMPA are indistinguishable. The XST-XSVT map is less sharp than the aforementioned methods, but this is already expected due to the method intrinsic lower lateral resolution. The results with OF, despite following closely the theoretical profile, present some artefacts/texture that is not observed with the previous methods - this texture is not captured by the NRMSE. The LCS results are not good towards the edge of the lens, where both the displacements and absorption are stronger.

Moving to the 500~µm radius lens (Fig.~\ref{fig:Comparison_lens_R500}), we see a better agreement between theoretical and experimental values. Here the pixel displacements are confined to $\pm$~1 pixel. Much like for the previous lens, XSVT, WXSVT and UMPA are indistinguishable. The XST-XSVT map remains the least sharp method but still presents excellent qualitative and quantitative results. OF still presents some texture near the lens edge, but those are much less prominent. The LCS method behaves much better than for the 50~µm case, but it is still noisy towards the edge of the lens where phase shifts approach the 1 pixel mark. The NRMSE remains between 6.2~\% and 6.9~\% for all methods mainly due to the modelling of the lens edge. Explicit tracking methods, however, present slightly lower NRMSE vales within the lens active area.

The third measurement of this series is of a 5000~µm radius lens. This is a very weak phase element and the pixel displacements are below a third of a pixel for the lens geometric aperture - edge effects are rather strong compared to the gradient inside the lens geometric aperture and hence, the elevated NRMSE values in Fig.~\ref{fig:Comparison_lens_R5000}. For this experiment, the explicit tracking methods are rather robust and menage to not only follow the slope of the gradient, but also show a slightly lower noise than the LCS method (implicit tracking). The OF algorithm under-performs and does not follow the straight line of the slope as evidenced by the gradients horizontal cuts. The OF image has three noticeable artefacts: two discontinuities on the left side of the lens (singularities/vortexes) and a conic shape on the right side. While the first remain source of speculation, the latter has been also observed in Fig.~\ref{fig:Comparison_lens_R500} for the same method and most likely comes from the scintillator or the carbon cover of the x-ray imaging system, which shields it from stray light. The LCS algorithm performs well for this sample, despite higher noise levels in the 2D map. We remind the reader that the noise levels shown here ($\sim1\big/10$ of a pixel) represent about $\sim210$~nrad of angular sensitivity for this experimental setup.

\subsection{Mouse knee: qualitative evaluation of a detailed sample under high-coherence illumination}\label{sec:mouse_knee}

\begin{figure*}
	\includegraphics[width=1\textwidth]{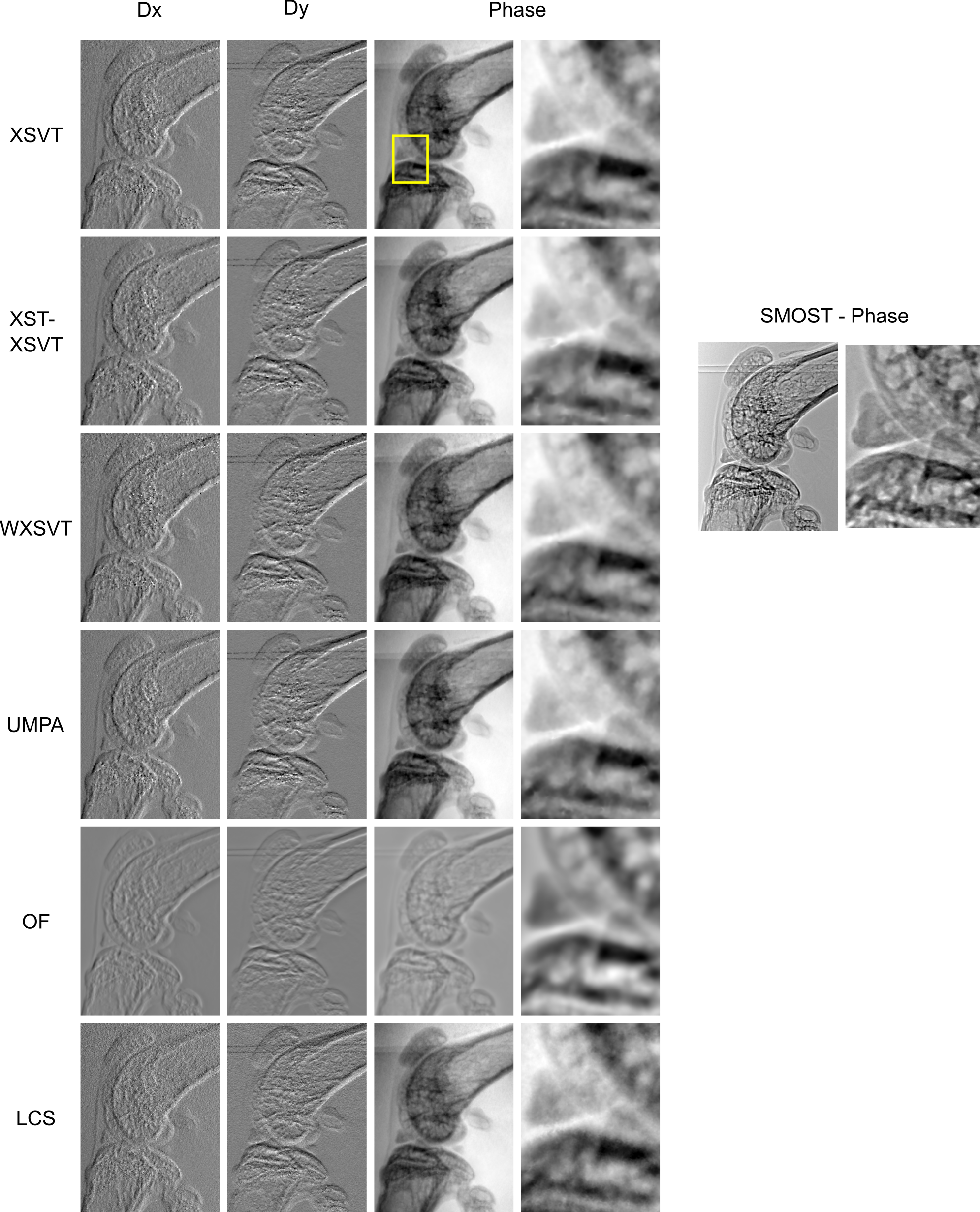}
	\caption{Displacement maps and phase images retrieved from acquisitions at 10 membrane positions with various phase retrieval algorithms.}
	\label{fig:Comparison_Mouse_knee_10pts}
\end{figure*}

The third study focused on a biological sample (mouse knee) under high-coherence illumination. In the previous sections, the ability to retrieve quantitative information was studied. Here we focus on image quality as it is difficult and rather impractical to obtain the theoretical phase shifts impinged for those kind of samples. The SMOST algorithm is now placed in the pool of tested SBI methods. Since the SMOST method obtains the phase rather than the phase-gradients like the remaining methods do, the Frankot Chellappa's integration method (FC) presented in \S\ref{sec:integration} will be applied. The results for 10 image pairs are presented in Fig. \ref{fig:Comparison_Mouse_knee_10pts}. In order to facilitate comparison, the gradient maps are all plotted with the same colour-scale. The phase maps are also plotted within the same scale, with the exception of the zoomed areas, where the best range for each subset is chosen to enhance small features and details.

All methods converge and retrieve the phase in a satisfactory way. The OF method, however, performs worse than the other algorithms and gives smooth/blurred images with a reduced dynamic range that makes some details difficult to see. This smoothing can be reduced by tuning the high-pass filter in the Fourier domain but this leads to the appearance of other artefacts. OF assumes a non-absorbing sample, a hypothesis which does not hold with highly absorbing samples such as bones. Since the absorption signal cannot be fully removed from the dataset, the signal retrieved by OF contains the filtered absorption. Isolating only the phase signal of this sample is not possible with this algorithm. The SMOST phase image is the one showing the most details and qualitatively, is the best reconstruction. However, it also shows interference fringes and the image is similar to the one obtained with a simple propagation-based phase imaging setup. This is due to the nature of the retrieved signal that, assuming a single material, combines information from the phase and attenuation. SMOST equation is similar to the PBI formulation, thus raising doubts about the role of membrane modulation in the image quality. 

\begin{figure*}
	\includegraphics[width=1\textwidth]{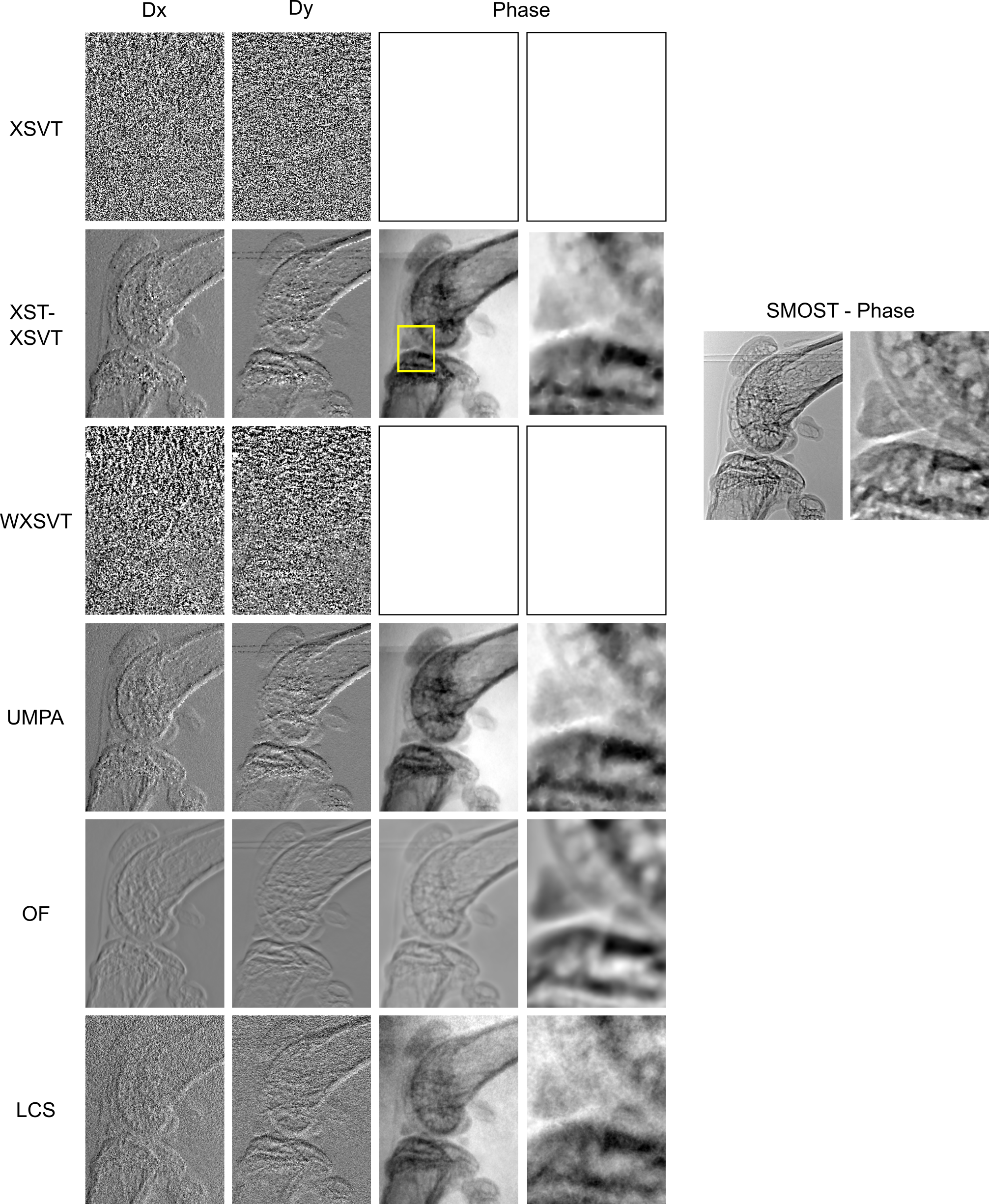}
	\caption{Displacement maps and phase images retrieved from acquisitions at 4 membrane positions with various phase retrieval algorithms. XSVT and WXSV fail in extracting the displacement maps, therefore the associated phase images are not shown}
	\label{fig:Comparison_Mouse_knee_4pts}
\end{figure*}

The reconstructions are now performed on a smaller data-set containing only 4 image pairs -- see Fig. \ref{fig:Comparison_Mouse_knee_4pts}. This is done to test the algorithms' robustness when reducing the samples exposure to x-rays and therefore reducing the delivered dose, which is relevant for medical imaging applications. Right away, we see that both XSVT and WXSVT fail. This can be easily explained by the fact that these algorithms track displacements through cross-correlation of 1D arrays: vectors of only four values are too small to meaningfully compute the cross-correlation. In the absence of a reliable displacement field, the integrated phase-shift images are not shown. For the XST-XSVT, UMPA and LCS, we observe a small loss of sensitivity and slightly less sharp edges: this is expected with fewer image pairs. The displacement field is nonetheless efficiently retrieved in areas where the speckle modulation and its gradient are high, albeit with a degraded quality when compared against the full data-set. Both OF and SMOST do not show a decrease in image quality: OF still underperforms, while SMOST retains the best reconstruction even with few image pairs.

\subsection{Domestic fly: qualitative evaluation of a detailed sample under low-coherence illumination}\label{sec:fly}

\begin{figure*}
	\includegraphics[width=1\textwidth]{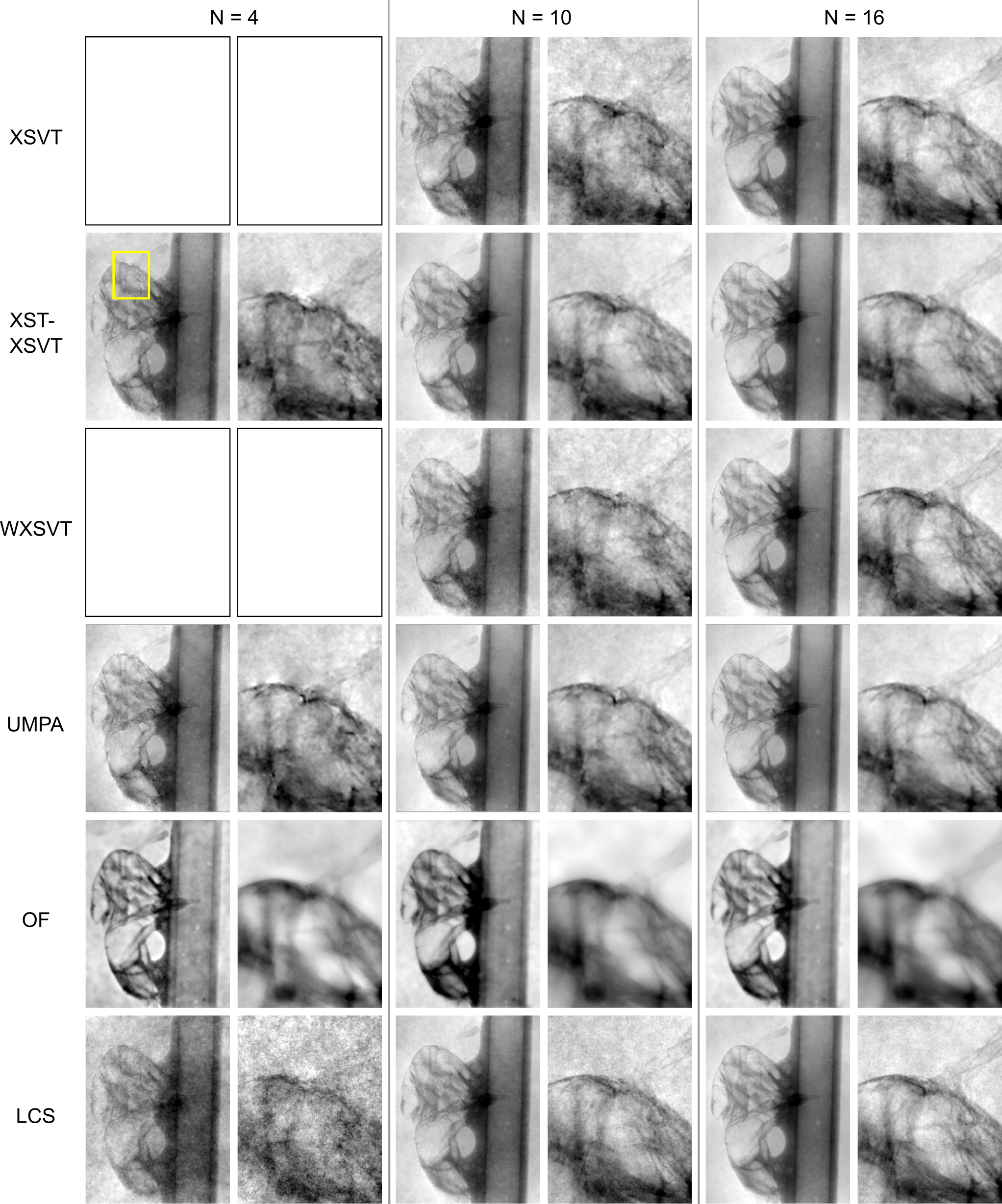}
	\caption{Phase images of a fly retrieved from a conventional set-up set of acquisitions with various available algorithms: XSVT, XST-XSVT, WXSVT, UMPA, OF, LCS from 4, 10 and 16 membrane positions.}
	\label{fig:ComparisonFly}
\end{figure*}

Finally, the algorithms were compared in the case of a low-coherence system, by using a laboratory source with polychromatic illumination (a tungsten anode operating at 40 kVp). This system also has a lower lateral resolution – the detector has a limited pixel size of 48 µm. To counteract the lower lateral resolution, an unsupervised Wiener deconvolution was applied to the acquisitions prior to the phase retrieval, to compensate for the blurring induced by the detector point spread function. For this comparison, 4, 10 and 16 membrane positions  were used. The resulting phase images are displayed in Fig. \ref{fig:ComparisonFly}. The structure of the noise varies significantly between the different algorithms. The large number of discontinuities in the displacement field, caused by the high spatial frequency content of the image, produces large variations in the integrated phase shift. For these reasons, an efficient image quality metric could not be found and the quality assessment remained qualitative. For the same reason, the grey level window was optimised for each algorithm. The results can be divided in three categories as follows: 
first, XST-XSVT, UMPA and LCS appear to give equivalent results. The UMPA phase appears slightly better with 4 positions but the LCS gives sharper lines at 16 positions;
second, OF gives comparable results independent of the number of positions, as observed for the mouse knee example. The images are blurred compared to the other algorithms, but in this case, the blur is balanced with an edge enhancement effect due to the Fourier frequency filter. It appears to be the least noisy result, however, the nature of the signal retrieved might be mainly due to filtered attenuation as previously discussed; third, the XSVT and WXSVT algorithms are unable to extract phase information from only 4 membrane positions but produce results compatible with the other algorithms already with 10 points. Considering that they use vectorial information over the number of positions and not windows as the other explicit algorithms, this is to be expected.

\begin{table*}[t]\caption{\label{tab:ComparisonBilan}Summary of the characteristics of the various phase retrieval algorithms.}
\resizebox{\textwidth}{!}{%
\begin{tabular}{rccccccc}
\hline \hline
 & \multicolumn{4}{c|}{\textbf{explicit tracking}} & \multicolumn{3}{c}{\textbf{implicit tracking}} \\ \cline{2-8} 
 &~~~XSVT~~~ & ~~~XST-XSVT~~~ & ~~~WXSVT~~~ & \multicolumn{1}{c|}{~~~UMPA~~~} & OF & SMOST & LCS \\ \cline{2-8} 
section (\S): & \ref{sec:XSVT} & \ref{sec:XST-XSVT}  & \ref{sec:WXSVT} & \multicolumn{1}{c|}{\ref{sec:UMPA}} & \ref{sec:OF} & \ref{sec:SMOST}  & \ref{sec:LCS}\\ \hline
quantitative: & yes & yes & yes & \multicolumn{1}{c|}{yes} & no &~~~\makecell{no due to filtering\\in the Fourier domain}~~~&~~~\makecell{yes for subpixel\\displacements}~~~\\ \hline
\makecell{sensitivity to \textbf{small}\\pixel displacements:} & + & + & + &  \multicolumn{1}{c|}{+} & / & / & ++ \\ \hline
\makecell{sensitivity to \textbf{large}\\pixel displacements:} & + & + & + & \multicolumn{1}{c|}{+} & / & / & -- \\ \hline
\makecell{robust to \textbf{decrease} in \\ membrane positions:} & - & + & - &  \multicolumn{1}{c|}{+} & ++ & ++ & $\pm (>3$) \\ \hline
\textbf{low} coherence & + & + & + &  \multicolumn{1}{c|}{+} & + & / & + \\ \hline
computational cost: & \makecell{high}  & very high & high &\multicolumn{1}{c|}{very high}&very low& very low & low \\ \hline \hline
\end{tabular}%
}
\end{table*}

\section{Conclusions and perspectives}
Table \ref{tab:ComparisonBilan} summarises the results obtained with the different algorithms. These methods present different advantages and drawbacks as discussed in the previous sections. If searching for high displacement quantitative measures, the XSVT, XST-XSVT, WXSVT and UMPA are the best options. The WXSVT has the advantage of a lower computational cost if fewer wavelet coefficients are used. For small displacements, the LCS gives quantitative results and might be the fastest computation option based on the resolution method. For synchrotron high-resolution imaging, XSVT, XST-XSVT, WXSVT, LCS and UMPA give similar results when a large number of membrane positions are acquired. With smaller numbers of points, XSVT and WXSVT are no longer good candidates. For conventional micro-focus source imaging of complex samples, the LCS, UMPA and XST-XSVT appear to give the best results that are quite similar to each other. Once again, for faster computation, the LCS is the best solution. OF and SMOST give interesting results when looking at complex samples but the nature of the extracted signal remains an issue. Further tests should be done to completely identify and characterise it. For example, the same SBI experiment could be performed with and without the modulator, to study the role of the membrane for signal extraction. A measurement without membrane would moreover provide a reliable measurement of the absorption, which could be decoupled from the phase signal.

\begin{acknowledgments}
We acknowledge the ESRF for providing beamtime at the BM05 and ID17 beamlines. The authors would like to thank Thomas Roth (XOG/ESRF), Sebastien Berujon (former BM05/ESRF), Jean-Pierre Guigay (ESRF), Jayde Livingstone (UGA) and Sylvain Bohic (INSERM/UGA) for valuable discussions. Part of this work was funded by LABEX PRIMES (ANR-11-LABX-0063) of Université de Lyon, within the program ``\textit{Investissements d’Avenir}'' (ANR-11-IDEX-0007)  and by the French National Research Agency (ANR) grant BREAKTHRU (ANR18-CE19-0003). E.B. would like to acknowledge the financial support of Inserm IRP project LINX.
\end{acknowledgments}

\newpage

\bibliography{Biblio}

\end{document}